\documentclass{emulateapj}

\shorttitle{Using gravitational-wave standard sirens}
\shortauthors{Holz \& Hughes}

\begin{document}

\title{Using gravitational-wave standard sirens}

\author{Daniel E. Holz}
\affil{Theoretical Division, Los Alamos National Laboratory,
Los Alamos, NM 87545 and\\
Kavli Institute for Cosmological
Physics and Department of Astronomy \& Astrophysics,\\ 
The University of Chicago, Chicago, IL 60637}
\and
\author{Scott A. Hughes}
\affil{Department of Physics and Center for Space Research,
Massachusetts Institute of Technology, 77 Massachusetts Avenue,
Cambridge, MA 02139}


\begin{abstract}
Gravitational waves (GWs) from supermassive binary black hole (BBH)
inspirals are potentially powerful standard sirens (the GW analog to
standard candles) \citep{schutz_nature,schutz_lh}. Because these
systems are well-modeled, the space-based GW observatory {\it LISA}
will be able to measure the luminosity distance (but not the redshift)
to some distant massive BBH systems with $1$--$10\%$ accuracy.  This
accuracy is largely limited by pointing error: GW sources generally
are poorly localized on the sky.  Localizing the binary independently
(e.g., through association with an electromagnetic counterpart)
greatly reduces this positional error.  An electromagnetic counterpart
may also allow determination of the event's redshift. In this case,
BBH coalescence would constitute an extremely precise (better than
1\%) standard candle visible to high redshift.  In practice,
gravitational lensing degrades this precision, though the candle
remains precise enough to provide useful information about the
distance-redshift relation.  Even if very rare, these GW standard
sirens would complement, and increase confidence in, other standard
candles.
\end{abstract}



\section{Introduction}
\label{sec:intro}

One of the major challenges to cosmology for the foreseeable future is
to understand ``dark energy'', the mysterious component responsible
for the apparent accelerating expansion of our Universe
{\citep{riess1998,perl1999,tonry03,knop03}}.  Dark energy can be
parameterized by its contribution to the universe's energy density,
$\Omega_X$, and its equation-of-state ratio, $w(z)$.  Of particular
interest will be measurements that probe $w(z)$, testing whether the
dark energy is a true cosmological constant [$w(z) = -1$] or whether
it arises, for example, from an evolving field
{\citep[e.g.][]{cds1998,ams2000}}.

One of our best observational probes of the dark energy is the
distance-redshift relation, which maps the expansion history of the
universe.  Much of our knowledge of this relation comes from
observations of distant Type Ia supernovae (SNe).  These SNe serve as
standard candles: their observed intensity can be calibrated to tell
us their luminosity distance, $D_L$
{\citep{phillips1993,rpk1995,wgap03}}.  As the redshift of a SN (or
its host) can also be measured, each SN puts a point on the
distance-redshift curve.  Future surveys (e.g., {\it
SuperNova/Acceleration Probe}\footnote{\url{http://snap.lbl.gov}},
{\it Large-aperture Synoptic Survey
Telescope}\footnote{\url{http://www.lssto.org}}) are expected to
measure thousands of Type Ia SNe, mapping the distance-redshift curve
over a large span of redshift with good statistical significance.

Type Ia SNe are excellent standard candles, with a (calibrated) peak
brightness thought to be known to about $15\%$.  A possible objection
to SNe as standard candles is the absence of a solid theoretical
underpinning.  Of particular concern is the possibility of evolution
in SN brightnesses, leading to unknown systematic errors
{\citep{dlw00}}.  In this article, we discuss a completely independent
standard candle: the gravitational-wave (GW) driven inspiral of
massive binary black holes (BBHs).  As GW detections can be thought of
as aural rather than optical~\citep{hughes2003}, a more appropriate
term for a GW standard candle is a ``standard siren''.\footnote{We
thank Sterl Phinney and Sean Carroll for suggesting this term.}
Because BBH systems are relatively simple and well modeled (at least
in the early ``inspiral'' phase of their coalescence), the GWs they
generate determine the source's luminosity distance with high
accuracy: typically $\delta D_L/D_L \sim 1$--$10\%$, with most of the
uncertainty arising from correlations with pointing errors
{\citep{hughes2002}}.  BBH merger events will follow the mergers of
galaxies and pregalactic structures at high redshift \citep{vhm2003}.
Though the merger rate is poorly understood, {\it LISA} is expected to
measure at least several events over its mission, especially as it is
sensitive to these waves to enormous distances
{\citep{richstone1998,haehnelt1998}}.

Since GWs do not provide the redshift of the source, BBH GW
measurements alone do not probe the distance-redshift relation.
However, as first noted by Bernard Schutz, should some kind of
``electromagnetic'' (EM) counterpart to a BBH GW event be identified,
the situation changes drastically {\citep{schutz_nature,schutz_lh}}.
First, by determining the source position, many correlations which set
the distance error are broken.  The error then drops immensely ---
below $0.5$--$1\%$ in many cases.  Second, a counterpart could
determine the source's redshift.  A BBH GW source coupled with an EM
counterpart could therefore constitute an exceedingly good standard
siren\footnote{It may also be possible to use the {\it distributions}
of observed binaries for cosmology, obviating the need for an EM
counterpart {\citep{cf93,finn96,wt97}}.  Unless the event rate is much
higher than currently expected, however, the statistical errors
associated with these distributions suggest that these methods will
not achieve accuracy sufficient to measure properties of the dark
energy equation of state, our primary focus.  Certainly other
cosmologically interesting measurements could be made.}.  We comment
at this point that, to date, there has not been a great deal of
careful analysis regarding the nature of EM counterparts which may
accompany a GW event.  We discuss briefly some ideas that have been
presented to date regarding the form that counterparts may take in
\S{\ref{sec:counterpart}} and \S\ref{sec:counterpart2}.  We hope that
the promise of this high quality candle will motivate additional
thinking on this issue.

In practice, gravitational lensing will limit the quality of this
candle.  GWs are lensed by intervening matter exactly as
electromagnetic waves are lensed {\citep{markovic93,wst96, tn03}}.  As
the waves propagate through our inhomogeneous universe, they are
magnified (or demagnified), inducing some error in our inferred
luminosity distance to the source.  As we discuss in
\S\ref{sec:lensing}, the distribution of errors is such that a BBH
candle will most likely be comparable in quality to a Type Ia SN
standard candle.  However --- and we strongly emphasize this point ---
the BBH candle will have entirely different systematics from SNe.
Concordance between the two types of measurement could thus alleviate
concerns about evolutionary effects in Type Ia SNe, and greatly
increase one's confidence in all standard candles.

\section{Distance determination with BBH GWs and {\it LISA}}
\label{sec:BBH_LISA}

Massive BBH coalescences are among the most luminous events in the
universe.  That luminosity (peaked at $\sim 10^{57}\,\mbox{erg/sec}$)
is radiated in GWs, which couple very weakly to matter.  The planned
space-based GW detector {\it LISA} (the {\it Laser Interferometer
Space Antenna}) will be sensitive to these BBH waves in the frequency
band $(10^{-5}$--$10^{-4})\,\mbox{Hz}\lesssim f\lesssim
0.1\,\mbox{Hz}$, making possible measurements from binaries with total
masses $m_1 + m_2\sim 10^3$--$10^6\,M_\odot$ {\citep{lisa}} out to
redshifts of at least $z\sim 5$--$10$ and possibly beyond
{\citep{hughes2002,v2004}}.  In this section we discuss how {\it LISA}
measurements determine the distance to a source, summarizing our model
of the waveform and {\it LISA}'s sensitivity and response, and
discussing the measurement precision we expect from measuring merging
black hole populations.

\subsection{Merging black hole GWs}

For this paper, the most interesting epoch of BBH coalescence is the
{\it inspiral}, when the binary's members are widely separated and
slowly spiral together due to backreaction from GW emission.  The GWs
from this epoch are well modeled using the post-Newtonian
approximation, roughly speaking an expansion in inverse separation of
a binary's members; see {\citet{b2002}} and references therein for
more detailed discussion.  We will not discuss waves from the {\it
merger} (in which the holes come into contact, forming a single body),
nor from the {\it ringdown} (the final, simple stage of the ringdown,
in which the merged binary is well-modeled as a single, distorted
black hole), as they do not substantially impact distance
determination.

Inspiral GWs encode the luminosity distance to a binary, its position
on the sky, its orientation, and information about certain
combinations of masses and spins; see {\citet{abiq2004}} and
{\citet{bdei2004}} for up-to-date discussion and details.  The
inspiral does {\it not} encode a source's cosmological redshift.
Redshift is instead entangled with the binary's evolution.  For
example, the masses $(m_1, m_2)$ impact orbital evolution as
timescales $(Gm_1/c^3, Gm_2/c^3)$.  These timescales redshift, so the
measured masses redshift: a binary with masses $(m_1, m_2)$ at
redshift $z$ is indistinguishible from a local binary with masses $[(1
+ z)m_1, (1 + z)m_2]$ (modulo amplitude).  This reflects the fact that
general relativity has no absolute scale.

In a reference frame centered on the solar system's barycenter, the
strongest harmonic of the inspiral GW's two polarizations has the form
\begin{eqnarray}
h_+ &=& {2{\cal M}_z^{5/3}[\pi f(t)]^{2/3}\over D_L}\left[1 + (\hat
L\cdot\hat n)^2\right] \cos[\Phi(t)]\;,
\\
h_\times &=& {4{\cal M}_z^{5/3}[\pi f(t)]^{2/3}(\hat L\cdot\hat
n)\over D_L} \sin[\Phi(t)]\;.
\label{eq:wave_bary}
\end{eqnarray}
The mass parameter ${\cal M}_z = (1 + z)(m_1 m_2)^{3/5}/(m_1 +
m_2)^{1/5}$ is the binary's redshifted ``chirp mass'', so called
because it largely sets the rate at which the binary's members spiral
towards one another, determining the ``chirp'' of the orbital
frequency.  The phase $\Phi(t)$ depends on intrinsic binary parameters
--- the masses and spins of its members {\citep[e.g.,][]{pw1995}}.  It
depends particularly strongly on ${\cal M}_z$; as a consequence, phase
coherent measurements of the waves will determine the chirp mass with
great precision {\citep{fc93,cf94}}.  The wave frequency $f(t) =
(1/2\pi)d\Phi/dt$.  The unit vector $\hat n$ points from the center of
the barycenter frame to the system, and hence defines its position on
the sky; $\hat L$ points along the binary's orbital angular momentum,
and hence defines its orientation.  Notice that the luminosity
distance $D_L$ appears in combination with these two angular factors.
Determining $D_L$ thus requires fixing these angles.  As we now
discuss, {\it LISA} is able to do so by virtue of its orbital motion.

\subsection{Merger GWs as measured by {\it LISA}}

\begin{figure}
\epsscale{1}
\plotone{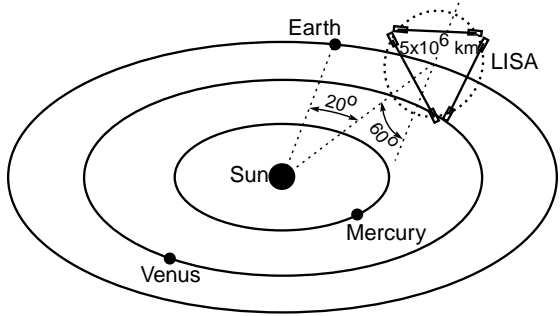}
\caption{Illustration of the {\it LISA} antenna's orbit.  The
constellation ``rolls'' as its centroid orbits the sun, completing one
full revolution for each orbit.}
\label{fig:lisa_orb}
\end{figure}

The {\it LISA} antenna consists of three spacecraft, arranged in
orbits about the Sun such that they form an equilateral triangle
(roughly; the armlengths are in general not equal, and in fact
oscillate --- albeit with periods much longer than that of the GWs we
aim to observe).  This triangle ``rolls'' as the spacecraft move
through their individual orbits, preserving the triangular formation.
The centroid of the constellation shares Earth's orbit, lagging by
$20^\circ$, so that it takes 1 year to orbit the Sun.  Figure
{\ref{fig:lisa_orb}} shows a schematic of the orbital configuration.
See {\citet{lisa}} for detailed discussion of the {\it LISA} mission
and its orbital configuration.

At least in the low frequency limit ($f < c/L$, where $L$ is
armlength), {\it LISA} can very usefully be regarded as two GW
detectors: the time varying armlength data, $(\delta L_1, \delta L_2,
\delta L_3)$ can be used to synthesize outputs for two equivalent
``L''-shaped detectors, with $90^\circ$ arms.  These ``equivalent
detectors'' are rotated by $45^\circ$ with respect to one another; see
{\citet{c98}} for details and derivation of this viewpoint.  The
datastream $s_{\rm I,II}$ of the these two equivalent detectors is
given by a weighted sum of the two GW polarizations, plus noise:
\begin{equation}
s_{\rm I,II}(t) = \frac{\sqrt{3}}{2}\left[F_{\rm I,II}^+(t) h_+(t)
+ F_{\rm I,II}^\times h_\times(t)\right] + n_{\rm I,II}(t)\;.
\label{eq:datastream}
\end{equation}
The prefactor $\sqrt{3}/2$ in this expression enters when converting
the ``real'' interferometer response to that of the synthesized
equivalent detectors.  The antenna functions $F^{+,\times}_{\rm I,II}$
depend on the orientation and position of the source relative to the
antenna.  Because of the antenna's orbital motion, the position and
orientation of the source relative to the antenna is continually
changing.  The motion of the detector thus modulates the measured
signal; the exact nature of the modulation depends upon the position
and orientation of the source.  We write the response functions as
time dependent functions to reflect this modulation.  Note also that
the waveform phasing is modified in an important manner by the
antenna's motion --- the orbital motion causes frequency as well as
amplitude modulation.  See {\citet{c98}} for further discussion.

We take the noises in the equivalent detectors, $n_{\rm I,II}(t)$, to
be uncorrelated, Gaussian random processes, with the same RMS values:
\begin{equation}
\langle n_{\rm I} n_{\rm II}\rangle = 0\;;\qquad
\langle n_{\rm I}^2 \rangle = \langle n_{\rm II}^2\rangle\;.
\end{equation}
In all of our analysis, we use the same noise model as that used by
{\citet{bc2004}} [their Eqs.\ (48)--(54)].  In our calculations, it is
necessary to introduce a low frequency cutoff --- a frequency at which
the sensitivity to GWs rapidly degrades.  This cutoff has important
implications for determining which binaries {\it LISA} can measure:
the frequency support of a binary's GW spectrum is inversely
proportional to its mass.  In other words, more massive binaries will
radiate at lower frequencies than less massive binaries.  The low
frequency cutoff thus determines the {\it maximum} binary black hole
mass accessible to {\it LISA} measurements.  It also determines the
amount of time for which a binary's waves are in band: a binary that
may only be in band for a few days when $f_{\rm low} = 10^{-4}\,{\rm
Hz}$ may be in band for many months when $f_{\rm low} = 3\times
10^{-5}\,{\rm Hz}$.

Unless stated otherwise, we have set $f_{\rm low} = 10^{-4}\,{\rm Hz}$
for the results we present here.  This is a somewhat conservative
choice; some members of the {\it LISA} mission design community
(particularly P.\ Bender) argue that {\it LISA} should have good
sensitivity down to frequencies $f \sim 10^{-5}\,{\rm Hz}$.
Accordingly, we have put $f_{\rm low} = 3\times10^{-5}\,{\rm Hz}$ in
several of our calculations.  We flag such cases when appropriate.

To understand more clearly how {\it LISA} extracts the luminosity
distance from measurements of a binary black hole inspiral, it is
useful to rewrite somewhat schematically the measured form of the
inspiral as follows:
\begin{equation}
h^{\rm meas}_{\rm I,II}(t) = {{\cal M}_z^{5/3}f(t)^{2/3}\over D_L}
{\cal F}_{\rm I,II}(\mbox{``angles''}, t)
\cos[\Phi(t) + \varphi_{\rm I,II}(\mbox{``angles''}, t)]\;.
\label{eq:wave_schem}
\end{equation}
We have subsumed the angle-dependent factors $(\hat L\cdot\hat n)$ and
$F^{+,\times}_{\rm I,II}$ into the schematic functions ${\cal F}_{\rm
I,II}(\mbox{``angles''}, t)$; we leave the dependence upon $t$ in
these functions as a reminder that the constellation's motion
modulates the waveform.  We have likewise written the phase
modulations imposed by the detector's response and motion in the
schematic form $\varphi_{\rm I,II}(\mbox{``angles''}, t)$.

From the form of equation\ (\ref{eq:wave_schem}) we see that the
luminosity distance is very strongly correlated with the redshifted
chirp mass, ${\cal M}_z$, and the various angles which set the
instantaneous waveform amplitude.  As already mentioned, ${\cal M}_z$
is typically determined with extremely high precision because it fixes
the phase evolution: typically, $\delta{\cal M}_z/ {\cal M}_z \lesssim
0.01\%$.  See {\citet{hughes2002}}, {\citet{v2004}} for examples
specific to {\it LISA}.

The modulations induced by {\it LISA}'s orbital motion make it
possible to measure sky position for events which last for at least a
fair fraction of {\it LISA}'s orbit.  We estimate the accuracy with
which position (among other parameters) is determined using a maximum
likelihood parameter estimation formalism {\citep{finn92}}: from the
detector's response to a given gravitational wave, we construct the
variance-covariance matrix $\Sigma^{ab}$.  Diagonal elements of this
matrix represent the rms error $\langle(\delta\chi^a)^2\rangle$ in a
source parameter $\chi^a$; off-diagonal components describe the degree
to which errors in parameters $\chi^a$ and $\chi^b$ are correlated.
See {\citet{hughes2002}} for discussion specifically tailored to this
application.

\subsection{Measurement accuracy distributions}
\label{sec:accuracy_dist}

\begin{figure}
\epsscale{1}
\plotone{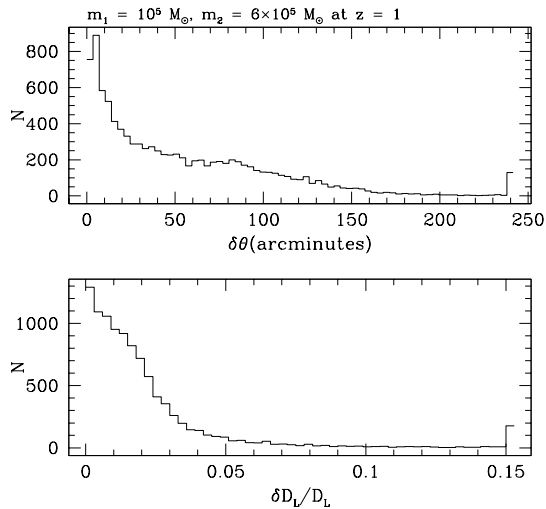}
\caption{Pointing and distance error distributions for measurements at
$z = 1$ of a binary of masses $m_1 = 10^5\,M_\odot$, $m_2 =
6\times10^5\,M_\odot$.  These distributions were made by Monte-Carlo
simulations of 10,000 {\it LISA} BBH measurements, randomly
distributing the binaries' positions, orientations, and merger times;
see {\citet{hughes2002}} for details.  The top distribution shows that
the most likely position error boxes have sides $\delta\theta\lesssim
10\,\mbox{arcminutes}$, spreading out to $\delta\theta\gtrsim
3^\circ$.  The distance distribution peaks at $\delta D_L/D_L \lesssim
1\%$, with most of the distribution confined to $\delta
D_L/D_L\lesssim 5\%$.}
\label{fig:z=1_m1=1e5_m2=6e5}
\end{figure}

To assess distance and position accuracy, we have estimated the
accuracy with which these parameters are measured for a wide range of
binary masses.  For each set of masses, we randomly distribute the sky
position and orientation of 10,000 such binaries.  We then calculate
the fractional accuracy with which distance is determined for each
binary, $\delta D_L/D_L$, as well as the angular sky position error
$\delta\theta$.  Figure {\ref{fig:z=1_m1=1e5_m2=6e5}} shows the
distribution we find in these quantities for binaries with $m_1 =
10^5\,M_\odot$, $m_2 = 6\times10^5\,M_\odot$ at $z = 1$.  We find that
the typical position determination is relatively poor --- these
binaries are fixed to an error box that, at best, is $\sim 5$
arcminutes on a side.  In most cases, the resolution is substantially
worse.  The distance determination, by contrast, is quite good: half
of these events have their distance determined with precision $\delta
D_L/D_L\lesssim 1\%$.

Table 1 summarizes the parameter determination distributions we find
for a wide range of masses.  For $\delta D_L/D_L$ and $\delta\theta$,
we give the 5\%, 25\%, 50\%, and 90\% likelihood values from the
distributions predicted by our Monte-Carlo calculation.  For example,
for binaries with $m_1 = 3\times10^4\,M_\odot$, $m_2 = 10^5\,M_\odot$
at $z = 1$, 25\% of all events have $\delta D_L/D_L \lesssim 0.006$
and localize the source to $\delta\theta \lesssim 14.6$ arcminutes;
90\% of all events with these masses and redshifts have $\delta
D_L/D_L \lesssim 0.029$ and localize the source to $\delta\theta
\lesssim 120$ arcminutes.

Notice that, in this table, the best pointing and distance
determination occurs for binaries that have a total (redshifted) mass
$(1 + z)(m_1 + m_2) \simeq \mbox{several}\times 10^5\,M_\odot$.  Two
competing effects drive this behavior.  First, for small binaries the
amplitude of the GWs is smaller; the degradation of their parameter
determination is due to reduced signal-to-noise ratio.  Larger
binaries are more interesting.  When such binaries enter {\it LISA}'s
sensitive band, they are closer to their final merger --- much less
inspiral remains once {\it LISA} begins measuring their waves.  They
therefore do not exhibit as many cycles of detector-motion-induced
modulation, and so their position angles are not as well determined.
In particular, we find that distance and position determination
rapidly degrades as binaries are made more massive than $(1+z)(m_1 +
m_2) \gtrsim \mbox{a few}\times 10^6\,M_\odot$.

The poor parameter determination tendency of large binaries can be
repaired somewhat by improving {\it LISA}'s low frequency sensitivity.
If the antenna has good sensitivity at lower frequencies, the span of
data containing good information about the inspiral can be lengthened.
Table 2 shows how well we measure distance and position when $f_{\rm
low}$ is reduced from $10^{-4}\,{\rm Hz}$ to $3\times 10^{-5}\,{\rm
Hz}$.  We now find that the distance is determined very precisely for
binaries with total mass $\mbox{(several)}\times 10^6\,M_\odot$.  Sky
position error is no worse than that achieved at lower masses ---
$\delta\theta \lesssim 10$ arcminutes in the best cases, and is more
typically a factor of a few larger than this.

\begin{figure}
\epsscale{1}
\plotone{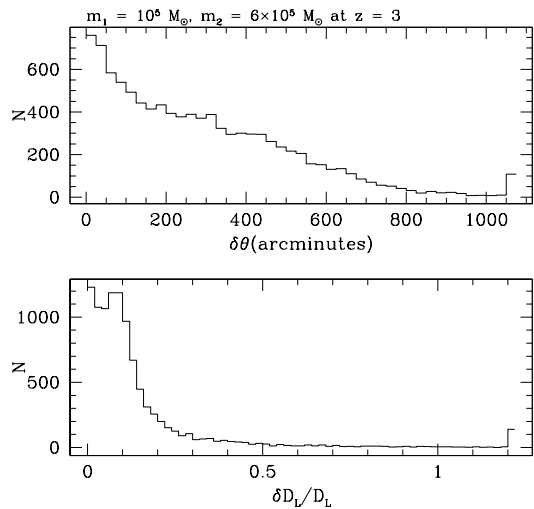}
\caption{Pointing and distance error distributions for measurements at
$z = 3$ of a binary of masses $m_1 = 10^5\,M_\odot$, $m_2 =
6\times10^5\,M_\odot$.  The distribution for position error is so
broad that we cannot really identify a ``most likely'' position error;
however, most of the distribution lies at $\delta\theta\lesssim
10^\circ$.  The distance distribution peaks at $\delta D_L/D_L
\lesssim 10\%$, with most of the distribution confined to $\delta
D_L/D_L\lesssim 30\%$.}
\label{fig:z=3_m1=1e5_m2=6e5}
\end{figure}

The same general story holds as we move to larger redshift.
Figure~{\ref{fig:z=3_m1=1e5_m2=6e5}} duplicates the content of
Figure~{\ref{fig:z=1_m1=1e5_m2=6e5}}, but with binaries at redshift $z
= 3$.  Likewise, Tables 3 and 4 duplicate the content of Tables 1 and
2, respectively, with all binaries placed at $z = 3$.  The overall
parameter determinations are worsened, as we would expect --- these
sources are much farther away, and so have greatly reduced
signal-to-noise.  In addition, the larger cosmological redshift shifts
the signal to lower frequencies, where much of it is lost in
low-frequency noise.  To quantify the impact of this effect, in Table
4 we present results showing what happens when we lower $f_{\rm low}$
from $10^{-4}\,{\rm Hz}$ to $3\times 10^{-5}\,{\rm Hz}$.  All cases
with $m_1 + m_2 \gtrsim 6\times 10^5\,M_\odot$ are substantially
improved by this fix.  Good low frequency performance will be
important for measuring high redshift binaries.  The best pointing
accuracy we find is $\delta\theta \lesssim 40$ arcminutes;
$\delta\theta \sim 1^\circ$ or larger is more typical.  The luminosity
distance can still be determined quite well --- we find errors of a
few percent in the best cases, and $\delta D_L/D_L \lesssim 15\%$ is
quite common.

It is worth emphasizing at this point that the results we present here
are most likely somewhat conservative.  By taking into account other
GW harmonics {\citep{mh2002}} and properly accounting for the
high-frequency structure of {\it LISA}'s response {\citep{seto}}, the
pointing accuracy, and thus distance accuracy, can be improved by a
factor of a few.  Properly accounting for modulations induced by
spin-orbit and spin-spin coupling can also improve pointing accuracy
and thus distance determination, in some cases significantly
{\citep{v2004}}.

Using a determination of $D_L$, we can {\it infer} the redshift by
using knowledge of the universe's geometry (the Hubble constant, mean
density of matter $\Omega_m$, and density of dark energy $\Omega_X$)
{\citep{hughes2002}}.  This makes possible interesting analyses (e.g.,
we can map the distribution of black hole masses as a function of
redshift), but presupposes rather than measures the distance-redshift
relation.

\section{The impact of a counterpart}
\label{sec:counterpart}

\begin{figure}
\epsscale{1}
\plotone{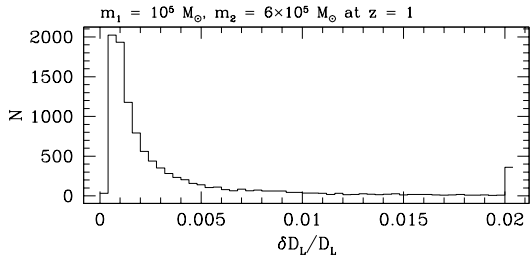}
\caption{Distance errors for BBH measurements at $z = 1$ with $m_1 =
10^5\,M_\odot$, $m_2 = 6\times10^5\,M_\odot$, assuming that an
electromagnetic counterpart allows precise sky position determination.
The peak error is at $\delta D_L/D_L \sim 0.1\%$, and is almost
entirely confined to $\delta D_L/D_L\lesssim 0.5\%$.}
\label{fig:pos_z=1_m1=1e5_m2=6e5}
\end{figure}

\begin{figure}
\epsscale{1}
\plotone{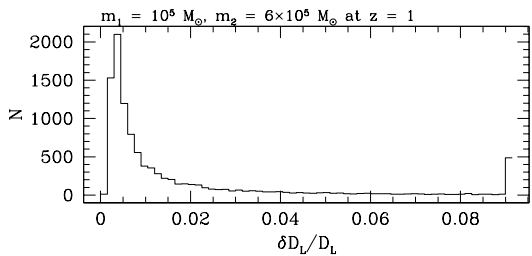}
\caption{Distance errors for BBH measurements at $z = 3$ with $m_1 =
10^5\,M_\odot$, $m_2 = 6\times10^5\,M_\odot$, assuming that an
electromagnetic counterpart allows precise sky position determination.
The peak error is at $\delta D_L/D_L \sim 0.5\%$, and is almost
entirely confined to $\delta D_L/D_L\lesssim 2\%$.}
\label{fig:pos_z=3_m1=1e5_m2=6e5}
\end{figure}

Parameter estimation improves dramatically when an EM counterpart to a
BBH GW event can be identified.  The counterpart will almost certainly
be pinpointed with far greater accuracy than is possible with GWs.
Correlations between position and distance are then broken, greatly
reducing the distance error.  An example of this improvement is shown
for $z = 1$ in Figure~{\ref{fig:pos_z=1_m1=1e5_m2=6e5}}.  The
distribution of distance errors peaks near $\delta D_L/D_L\sim 0.1\%$,
and is largely confined to $\delta D_L/D_L\lesssim 1\%$.  Similar
results are seen for $z = 3$ (Fig.\
{\ref{fig:pos_z=3_m1=1e5_m2=6e5}}), albeit with precision degraded by
a factor of a few due to lower signal-to-noise.  Comparing to the
lower panel of Figures~{\ref{fig:z=1_m1=1e5_m2=6e5}}
and~{\ref{fig:z=3_m1=1e5_m2=6e5}}, we see that associating the event
with a counterpart improves distance accuracy by roughly an order of
magnitude.  This rough level of improvement holds over a wide band of
mass and redshift.

\begin{figure}
\epsscale{1}
\plotone{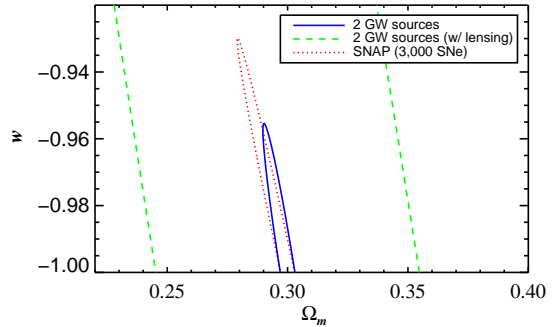}
\caption{Likelihood contours for measurement of the matter density
$\Omega_m$ and dark energy equation of state parameter $w$ (with the
pressure and density of the dark energy related by $p = w\rho$).  We
assume that the universe is flat, and that the underlying model has
$\Omega_m = 0.3$ and $w = -1$. The two GW sources are at $z = 1$ and
$z = 3$, while the SNAP SNe are evenly distributed within $0.7 < z <
1.7$.  }
\label{fig:contour}
\end{figure}

A correlated GW/EM measurement will be particularly important if the
counterpart provides a redshift as well as an improved $D_L$.  Such a
measurement would constitute a powerful standard candle, probing the
distance-redshift relation in a manner complementary to other candles,
such as Type Ia supernovae.  Assuming a flat universe and a Hubble
constant $h_0 = 0.65$, we ask how well the matter density $\Omega_m$
and dark energy equation-of-state ratio $w$ can be measured.  Figure\
{\ref{fig:contour}} shows estimated likelihood contours
($1$--$\sigma$) in the ($\Omega_m, w$) plane.  The dotted line shows
the contour expected for measurements of 3,000 SNe evenly distributed
within $0.7 < z < 1.7$ (reasonable choices for {\it SNAP}); the solid
contour is for two GW events, one each at $z = 1$ and $z = 3$.  (We
will discuss the dashed line further below.)  Redshift and distance
are measured with such accuracy that the contours are extremely tight
even for only a small number of sources.

We emphasize the current poor understanding of EM counterparts to BBH
GW events, although the possibility of such counterparts has been
discussed for quite some time (e.g., Begelman, Blandford, \& Rees
1980).  {\citet{mp04}} have recently examined the evolution of gas in
the environment of a merging binary, and show that there is likely to
be a delayed electromagnetic afterglow.  They find that the merging
binary carves a hollow region in the volume of circumbinary gas.  The
binary separation shrinks faster than the inner edge of the hollowed
region; thus, as the coalescence proceeds, there should be no
substantial accretion of material onto the system.  The gas falls onto
the merged remnant several years after the merger, leading to an
afterglow that should be measurable by next generation x-ray
telescopes.

Other models suggest that there may be an electromagnetic {\it
precursor} to the merger, rather than a delayed glow.  One example is
discussed by {\citet{an2002}}.  They argue that gas is driven onto the
larger member of the binary by the secondary's inspiral, leading to
super-Eddington accretion.  In this model, much of the inner disk may
be expelled from the system in a high velocity ($\sim
10^4\,\mbox{km/sec}$) outflow.  Such strong outflows could flag a
recent or impending merger.  A similar family of models
{\citep{s1988,lv1996}} explains periodic variations in the BL Lac
object OJ 287 by a tight, eccentric binary system with mass ratio of
about 1:100.  Flaring outbursts from this quasar are explained as
arising from the secondary periodic crossing of the primary's
accretion disk.  Given the great payoff that would follow from
associating a counterpart to a GW event, we strongly advocate
continuing to develop and refine models of BBH mergers.

It is worth noting that, for a small fraction of binaries (assuming a
sufficiently high event rate), {\it LISA} will provide an error box of
$\lesssim 5$ arcmin, and an estimate of the time of merger about a day
in advance.  Regardless of the state of theoretical predictions, we
imagine that in such cases there will be great interest in searching
the GW source error box for any observational counterparts to the
merger.  Indeed, as we briefly discuss in \S\ref{sec:counterpart2},
the number of relevant galaxies in the {\it LISA} error box may be
fairly small, and so associating an EM counterpart to the GW event may
be tractable.

\section{Gravitational lensing}
\label{sec:lensing}

Having discussed the impressive quality of GW standard sirens, we turn
now to an important caveat: the impact of gravitational lensing on the
distance measurement.  GWs are lensed exactly as EM radiation is
lensed.  Since we expect BBH events to come from rather large redshift
($z\gtrsim 1$), weak lensing in the GW datasets should be common
{\citep{markovic93,wst96}} (in addition to the occasional
strongly-lensed source).

\begin{figure}
\epsscale{1}
\plotone{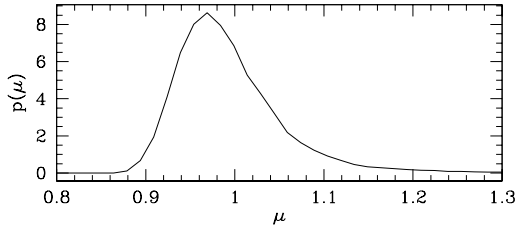}
\caption{The differential probability of magnification by
gravitational lensing, $p(\mu)$, for sources at $z = 1.5$ in a
concordance universe; see {\citet{whm2002}} for details.}
\label{fig:lens_dist}
\end{figure}

A lens with magnification $\mu$ will distort the inferred luminosity
distance to the source: if the true distance is $D_L$, we measure
$D_L/\sqrt{\mu}$, incurring a ``systematic'' error $\Delta D_L/D_L = 1
- 1/\sqrt{\mu}$.  We estimate the error such lensing is likely to
introduce by convolving this quantity with the expected magnification
distribution, $p(\mu)$ {\citep{hw98,whm2002}}; an example of this
distribution is shown in Figure~{\ref{fig:lens_dist}}.  Using
parameters appropriate to a $\Lambda$CDM model of the universe, we
find a mean error at $z = 2$ of $\langle\Delta D_L/D_L\rangle \simeq
0.005$, with a variance $\sqrt{\langle (\Delta D_L/D_L)^2\rangle}\simeq 0.09$
(approximating the lensing by a Gaussian; see \citet{hl05}).
The dashed line in Figure~{\ref{fig:contour}} shows the
contour we expect from the 2 GW sources when lensing errors are
included.  The parameter accuracies are significantly degraded.

Of course, this magnification bias affects {\it all} standard candles,
not just GWs.  The rate of Type Ia SNe, however, is high enough to
sufficiently sample the entire lensing distribution, and thus average
away the bias.  Missions such as {\it SNAP} are designed to observe
thousands of SNe at high redshift, in large part to overcome
gravitational lensing.  Indeed, this may allow one to measure the
lensing signal well enough to infer characteristics of the lensing
matter {\citep{ms99,sh99}}.  This is unlikely to be the case with BBH
GWs: the rate of mergers will likely be much lower than that of SNe
{\citep{richstone1998,haehnelt1998}}, so we cannot count on enormous
numbers of events.  We also emphasize that we do not expect to be able
to correct for gravitational lensing effects on a case-by-case basis
{\citep{dalal03}}.  Lensing, therefore, will introduce an
insurmountable error of $\sim5$--$10\%$ for each individual high
redshift event, significantly greater than the intrinsic distance
error.

\section{Identifying the counterpart}
\label{sec:counterpart2}

In order to provide data on the distance-redshift curve, a GW event
must be associated with an ``electromagnetic'' counterpart --- GWs
provide an accurate measure of luminosity distance, but give no direct
information about redshift.  This is the weakest link in our analysis:
we do not know whether such counterparts exist.  However, a simple
counting argument suggests that the number of relevant galaxies in the
{\it LISA} error cube may be fairly small.  We approximate the
redshift distribution of source galaxies by
\begin{equation}
\frac{dN}{dRd\Omega} \propto R^\alpha \exp\left[ -(R/R_*)^\beta\right
]\;,
\label{eq:galaxy_density}
\end{equation}
where $R$ is the comoving distance; we take $\alpha = 1$, $\beta = 4$,
and $R_* = c/H_0$ {\citep{kaiser92,hu99}}. We normalize this to a
projected number density of
\begin{equation}
\frac{dN}{d\Omega} = \int dR\,\frac{dN}{dRd\Omega} \simeq
300\ {\rm galaxies/arcmin}^2\;,
\end{equation}
approximating the Hubble Deep Field \citep{hdf96}.

\begin{figure}
\epsscale{1}
\plotone{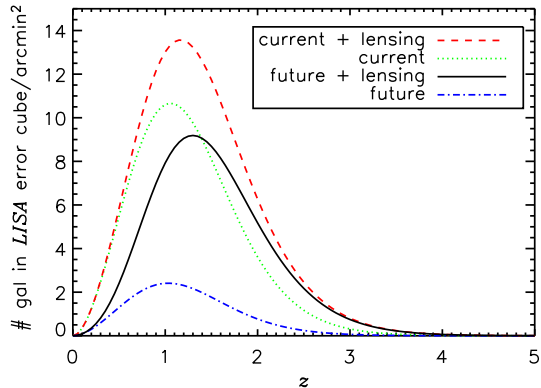}
\caption{Number of candidate host galaxies per square arcmin of {\it
LISA}\/ error cube for a supermassive binary black hole coalescence
event, as a function of redshift of the event. Dotted (green) line
utilizes current uncertainties in cosmological
parameters~\citep{wmap03}. Dash-dotted (blue) line represents possible
future improvements in these parameters (1\% in $\Omega_m$,
$\Omega_{\Lambda}$, $h_0$). Dashed (red) line includes the degradation
in the depth of the error cube due to gravitational lensing, for
current cosmological uncertainties. Solid (black) line represents
future cosmological uncertainties, with the inclusion of gravitational
lensing degradation. The 1 arcmin {\it LISA} sky position error
considered here is optimistic, though not implausible (see text). This
choice allows for a straightforward scaling of the curves to larger
positional errors.}
\label{fig:n_gal}
\end{figure}

As we have discussed extensively, a GW measurement of a binary black
hole merger determines the position on the sky to within some error
$\delta\theta$, and determines the luminosity distance to within some
error $\delta D_L$.  By assuming a cosmological model we can convert
the measured luminosity distance, and its error, to any other desired
cosmic distance measure.  Denoting by $\delta R(\delta D_L;\delta{\rm
cosmology})$ the error in comoving distance due to both the GW
measurement error and the uncertainty in cosmological parameters, the
number of galaxies which lie in the 3-dimensional GW error cube is
\begin{equation}
N_{\rm error\ cube} \simeq \frac{dN}{dRd\Omega}\times \delta\theta^2
\delta R(\delta D_L;\delta{\rm cosmology})\;.
\label{eq:Nrelevant}
\end{equation}
Figure~\ref{fig:n_gal} shows four realizations of $N_{\rm error\
cube}$ as a function of GW event redshift, $z$.  We have scaled to a
near best-case pointing accuracy of $\delta\theta = 1$ arcmin; we
emphasize that this is an optimistic, though not implausible, pointing
error. From the Tables it is apparent than about 5\% of binaries with
masses and redshifts in the {\it LISA} sweetspot have positional
errors $\delta\theta\lesssim5\ \mbox{arcmin}$. As discussed towards
the end of \S\ref{sec:accuracy_dist}, these results may be
conservative --- accounting for spin-induced precessional effects may
allow certain degeneracies to be broken and improve {\it LISA}'s
pointing accuracy {\citep{v2004}}.  [A fiducial pointing error of 1
arcminute also makes it very simple, by Eq.\ (\ref{eq:Nrelevant}), to
scale to larger values.]  The number of galaxies decreases
significantly as uncertainties in cosmological parameters are reduced
(as is to be expected by the time that {\it LISA} is operating).  In
Figure~\ref{fig:n_gal} we include the curve for the current state of
errors on cosmological parameters, as well as for expected future,
percent-level, measurements.  As was discussed in the previous
section, gravitational lensing adds a further, insurmountable error to
the GW measurement of $\delta D_L$.  We approximate the lensing
effects by a Gaussian in magnification, with variance given by
$\sigma_{\rm lensing}=0.088z$ \citep{hl05}. Although this expression
is strictly appropriate only for high source statistics (as otherwise
the lensing distribution is non-Gaussian), it is a sufficiently good
approximation for present purposes.  With the inclusion of lensing
errors in addition to cosmological parameter uncertainties, we find
$\lesssim 10$--$20$ potential counterpart galaxies in 1 arcmin of LISA
error cube.

The number of candidate host objects for a galactic binary black hole
merger is thus likely to be tractable.  What remains is to find a way
to identify which of the dozen or so candidate objects is in fact the
host of the merger.  Useful models exploring the signatures that may
constitute ``precursors'' of the merger exist
{\citep{an2002,s1988,lv1996}}; we are hopeful that such models can be
extended to the ranges of mass and mass ratio of binaries that are
likely to be observationally interesting to {\it LISA}.

The remnant of the merger is very likely to have an irregular
morphology. In addition, the host galaxy may be in an active phase.
{\citet{mp04}} have recently developed a model for the late x-ray
afterglow of a BBH merger. {\citet{kfmh05}} consider a scenario in
which a merger remnant is associated with a quasar, and argue that in
this case the paucity of quasars will make the identification of a
counterpart significantly easier.  In both of these models the merger
remnant ``lights up'', and is thus relatively easy to identify in the
positional error box.  Even if the remnant does not light up, other
information will help winnow the list of candidate host galaxies.  For
example, the GWs will measure the masses of the black holes with good
accuracy; by using properties such as the $M_{\rm BH}$--$\sigma$
relation we can estimate specific properties (e.g., kinematics,
luminosity) that the host galaxy would be expected to have.

Observations of the {\it LISA} error boxes will no doubt be
undertaken, regardless of the state of theoretical predictions. It is
only by such direct observations that we will determine whether or not
EM counterparts to BBH GW sources can be identified.

\section{Conclusions}
\label{sec:conclude}

Because of their potential as an independent set of standard candles,
BBH GW standard sirens can make an important contribution to programs
to map the distance-redshift relation over a large span of redshift.
Although the intrinsic precision of these candles is phenomenal, this
precision will be limited in practice because of gravitational
lensing.  With lensing taken into account, the accuracy of the BBH GW
candle is comparable to (or perhaps slightly better than) a Type Ia
SN.  It is sobering to note that we are already approaching the point
at which lensing, rather than intrinsic dispersion, limits our ability
to use standard candles.

We emphasize that the systematics of BBH events are entirely different
from those of Type Ia SNe.  As such, the greatest impact of BBH
standard sirens may be to verify, and thereby increase our confidence
in, other standard candles. The utility of these GW standard sirens
depends on the identification of an electromagnetic counterpart,
through which a redshift to the source can be determined. It is not
unlikely that, at least for some of the best-observed systems, a
counterpart will be found. If this is the case, the BBH GW source
would become an exceptionally precise standard siren.

\acknowledgments

We thank Zoltan Haiman, Kristen Menou, Steinn Sigurdsson, and
particularly Shane Larson for useful discussions; we also thank Sam
Finn for detailed and helpful comments on a previous version of this
manuscript.  We thank Sean Carroll and Sterl Phinney for independently
suggesting that the gravitational-wave analogue of the standard candle
be named the ``standard siren''.  We are extremely grateful to
Emanuele Berti for pointing out that the speed of the Monte-Carlo code
used in this analysis could be vastly improved by changing our
integration routines.  Finally, we thank the referee for suggesting
that morphological characteristics and the $M_{\rm BH}$--$\sigma$
relation may help to identify a merger event's host galaxy.  This work
was initiated when the authors were at the Kavli Institute for
Theoretical Physics (Santa Barbara), and were supported by NSF Grant
PHY-9907949.  DEH is supported by NSF Grant PHY-0114422, and
gratefully acknowledges a Feynman Fellowship from LANL. SAH is
supported by NASA Grant NAG5-12906 and NSF Grant PHY-0244424.

\clearpage

\begin{deluxetable}{cccc}
\tablecaption{Measurement precision at $z = 1$ with $f_{\rm low} =
10^{-4}\,{\rm Hz}$}
\tablehead{
\colhead{$m_1\,(M_\odot)$} &
\colhead{$m_2\,(M_\odot)$} &
\colhead{$\delta D_L/D_L$} &
\colhead{$\delta\theta$} \\
\colhead{} &
\colhead{} &
\colhead{(5\%, 25\%, 50\%, 90\%)} &
\colhead{(5\%, 25\%, 50\%, 90\%)} }
\startdata
$10^4$ & $10^4$ & (0.005, 0.010, 0.014, 0.042) &
(14.3, 28.3, 48.5, 117) arcmin \\
$10^4$ & $3\times10^4$ & (0.003, 0.008, 0.013, 0.037) &
(11.0, 22.1, 41.6, 111) arcmin \\
$10^4$ & $6\times10^4$ & (0.003, 0.007, 0.013, 0.036) &
(9.05, 19.1, 40.4, 109) arcmin \\
$10^4$ & $10^5$ & (0.005, 0.007, 0.013, 0.035) &
(7.85, 18.3, 39.5, 110) arcmin \\
$3\times10^4$ & $10^5$ & (0.002, 0.006, 0.013, 0.029) &
(5.26, 14.6, 37.4, 120) arcmin \\
$6\times10^4$ & $10^5$ & (0.002, 0.006, 0.012, 0.037) &
(4.07, 12.9, 35.0, 120) arcmin \\
$10^5$ & $10^5$ & (0.002, 0.005, 0.012, 0.034) &
(3.36, 11.7, 33.4, 117) arcmin \\
$10^5$ & $3\times10^5$ & (0.001, 0.005, 0.012, 0.035) &
(2.71, 10.6, 33.1, 116) arcmin \\
$10^5$ & $6\times10^5$ & (0.001, 0.006, 0.014, 0.044) &
(2.82, 12.0, 38.8, 120) arcmin \\
$10^5$ & $10^6$ & (0.002, 0.009, 0.017, 0.053) &
(3.89, 18.5, 50.9, 126) arcmin \\
$3\times10^5$ & $10^6$ & (0.002, 0.013, 0.026, 0.087) &
(4.65, 29.7, 71.0, 172) arcmin \\
$6\times10^5$ & $10^6$ & (0.003, 0.019, 0.035, 0.122) &
(5.60, 39.2, 93.6, 220) arcmin \\
$10^6$ & $10^6$ & (0.004, 0.024, 0.043, 0.149) &
(6.36, 52.2, 118, 271) arcmin \\
\enddata
\end{deluxetable}

\begin{deluxetable}{cccc}
\tablecaption{Measurement precision at $z = 1$ with $f_{\rm low} =
3\times10^{-5}\,{\rm Hz}$}
\tablehead{
\colhead{$m_1\,(M_\odot)$} &
\colhead{$m_2\,(M_\odot)$} &
\colhead{$\delta D_L/D_L$} &
\colhead{$\delta\theta$} \\
\colhead{} &
\colhead{} &
\colhead{(5\%, 25\%, 50\%, 90\%)} &
\colhead{(5\%, 25\%, 50\%, 90\%)} }
\startdata
$10^5$ & $10^5$ & (0.002, 0.005, 0.012, 0.036) &
(3.35, 11.4, 33.2, 117) arcmin \\
$10^5$ & $3\times10^5$ & (0.001, 0.005, 0.011, 0.034) &
(2.68, 10.4, 31.0, 107) arcmin \\
$10^5$ & $6\times10^5$ & (0.001, 0.005, 0.010, 0.032) &
(2.71, 9.85, 29.7, 103) arcmin \\
$10^5$ & $10^6$ & (0.001, 0.005, 0.010, 0.031) &
(3.27, 10.5, 29.1, 101) arcmin \\
$3\times10^5$ & $10^6$ & (0.001, 0.004, 0.009, 0.028) &
(2.45, 8.77, 25.8, 89.9) arcmin \\
$6\times10^5$ & $10^6$ & (0.001, 0.004, 0.009, 0.026) &
(2.25, 8.5, 24.8, 84.3) arcmin \\
$10^6$ & $10^6$ & (0.001, 0.004, 0.009, 0.026) &
(2.39, 9.05, 25.3, 83.5) arcmin \\
$10^6$ & $3\times10^6$ & (0.002, 0.006, 0.011, 0.034) &
(4.56, 12.6, 34.6, 93.8) arcmin \\
$10^6$ & $6\times10^6$ & (0.003, 0.008, 0.016, 0.051) &
(7.16, 20.6, 45.9, 110) arcmin \\
$10^6$ & $10^7$ & (0.003, 0.011, 0.021, 0.071) &
(9.20, 27.0, 57.1, 136) arcmin \\
\enddata
\end{deluxetable}

\begin{deluxetable}{cccc}
\tablecaption{Measurement precision at $z = 3$ with $f_{\rm low} =
10^{-4}\,{\rm Hz}$}
\tablehead{
\colhead{$m_1\,(M_\odot)$} &
\colhead{$m_2\,(M_\odot)$} &
\colhead{$\delta D_L/D_L$} &
\colhead{$\delta\theta$} \\
\colhead{} &
\colhead{} &
\colhead{(5\%, 25\%, 50\%, 90\%)} &
\colhead{(5\%, 25\%, 50\%, 90\%)} }
\startdata
$10^4$ & $10^4$ & (0.013, 0.029, 0.051, 0.143) &
(37.7, 78.2, 158, 422) arcmin \\
$10^4$ & $3\times10^4$ & (0.010, 0.026, 0.050, 0.135) &
(27.2, 66.8, 150, 428) arcmin \\
$10^4$ & $6\times10^4$ & (0.008, 0.024, 0.050, 0.145) &
(22.0, 57.7, 140, 442) arcmin \\
$10^4$ & $10^5$ & (0.008, 0.024, 0.050, 0.142) &
(19.3, 55.4, 143, 465) arcmin \\
$3\times10^4$ & $10^5$ & (0.006, 0.019, 0.044, 0.131) &
(12.6, 44.5, 125, 444) arcmin \\
$6\times10^4$ & $10^5$ & (0.006, 0.021, 0.044, 0.128) &
(10.1, 42.6, 132, 429) arcmin \\
$10^5$ & $10^5$ & (0.005, 0.024, 0.049, 0.141) &
(9.55, 44.8, 144, 430) arcmin \\
$10^5$ & $3\times10^5$ & (0.007, 0.034, 0.069, 0.213) &
(11.3, 69.9, 193, 485) arcmin \\
$10^5$ & $6\times10^5$ & (0.008, 0.044, 0.087, 0.287) &
(17.5, 96.1, 240, 593) arcmin \\
$10^5$ & $10^6$ & (0.009, 0.058, 0.111, 0.378) &
(23.4, 127, 304, 734) arcmin \\
\enddata
\end{deluxetable}

\begin{deluxetable}{cccc}
\tablecaption{Measurement precision at $z = 3$ with $f_{\rm low} =
3\times 10^{-5}\,{\rm Hz}$}
\tablehead{
\colhead{$m_1\,(M_\odot)$} &
\colhead{$m_2\,(M_\odot)$} &
\colhead{$\delta D_L/D_L$} &
\colhead{$\delta\theta$} \\
\colhead{} &
\colhead{} &
\colhead{(5\%, 25\%, 50\%, 90\%)} &
\colhead{(5\%, 25\%, 50\%, 90\%)} }
\startdata
$10^4$ & $10^4$ & (0.013, 0.029, 0.050, 0.136) &
(37.4, 76.0, 156, 423) arcmin \\
$10^4$ & $3\times10^4$ & (0.010, 0.026, 0.050, 0.137) &
(26.5, 62.7, 149, 437) arcmin \\
$10^4$ & $6\times10^4$ & (0.009, 0.024, 0.049, 0.140) &
(21.9, 57.8, 141, 441) arcmin \\
$10^4$ & $10^5$ & (0.007, 0.023, 0.049, 0.139) &
(19.0, 55.3, 143, 465) arcmin \\
$3\times10^4$ & $10^5$ & (0.006, 0.019, 0.043, 0.135) &
(12.4, 43.6, 124, 429) arcmin \\
$6\times10^4$ & $10^5$ & (0.005, 0.018, 0.040, 0.126) &
(10.4, 38.8, 114, 404) arcmin \\
$10^5$ & $10^5$ & (0.004, 0.016, 0.038, 0.120) &
(8.85, 34.7, 106, 384) arcmin \\
$10^5$ & $3\times10^5$ & (0.004, 0.016, 0.035, 0.108) &
(8.26, 32.4, 98.7, 360) arcmin \\
$10^5$ & $6\times10^5$ & (0.005, 0.016, 0.035, 0.105) &
(10.9, 36.0, 100, 346) arcmin \\
$10^5$ & $10^6$ & (0.006, 0.017, 0.035, 0.106) &
(15.2, 41.6, 105, 349) arcmin \\
$3\times10^5$ & $10^6$ & (0.005, 0.017, 0.034, 0.100) &
(12.2, 37.3, 101, 327) arcmin \\
$6\times10^5$ & $10^6$ & (0.006, 0.019, 0.039, 0.115) &
(13.1, 42.1, 120, 336) arcmin \\
$10^6$ & $10^6$ & (0.007, 0.025, 0.049, 0.146) &
(16.9, 55.5, 145, 359) arcmin \\
$10^6$ & $3\times10^6$ & (0.009, 0.040, 0.077, 0.240) &
(24.8, 89.1, 208, 507) arcmin \\
$10^6$ & $6\times10^6$ & (0.014, 0.063, 0.116, 0.392) &
(35.4, 142, 322, 757) arcmin \\
\enddata
\end{deluxetable}

\end{document}